\documentclass[reprint, superscriptaddress, amsmath, amssymb, aps]{revtex4-2}

\usepackage{graphicx}
\usepackage{dcolumn}
\usepackage{bm}
\usepackage{xcolor}
\usepackage{mathtools}
\usepackage{physics}
\usepackage[normalem]{ulem} 
\usepackage{lineno}

\usepackage{hyperref} 
\hypersetup{colorlinks=true, citecolor=blue}

\begin{document}

\title{Tuning of Strong Nonlinearity in rf SQUID Meta-Atoms}

\author{Ethan Zack}
 \email{ezack@terpmail.umd.edu}
 \affiliation{Quantum Materials Center, Department of Physics, University of Maryland, College Park, MD 20742-4111, USA}
 
\author{Daimeng Zhang}
 \affiliation{Quantum Materials Center, Department of Physics, University of Maryland, College Park, MD 20742-4111, USA}
 
 \author{Melissa Trepanier}
 \affiliation{Quantum Materials Center, Department of Physics, University of Maryland, College Park, MD 20742-4111, USA}
 
 \author{Jingnan Cai}
 \affiliation{Quantum Materials Center, Department of Physics, University of Maryland, College Park, MD 20742-4111, USA}
 
 \author{Tamin Tai}
 \affiliation{Quantum Materials Center, Department of Physics, University of Maryland, College Park, MD 20742-4111, USA}
 
\author{Nikos Lazarides}
 \affiliation{Department of Physics, University of Crete, 71003 Herakleio, Greece}
 
\author{Johanne Hizanidis}
\affiliation{Department of Physics, University of Crete, 71003 Herakleio, Greece}

\author{Steven M. Anlage}
 \email{anlage@umd.edu}
 \affiliation{Quantum Materials Center, Department of Physics, University of Maryland, College Park, MD 20742-4111, USA}

\date{\today}
 
\begin{abstract}

Strong nonlinearity of a self-resonant radio frequency superconducting quantum interference device (rf-SQUID) meta-atom is explored via intermodulation (IM) measurements. Previous work in zero dc magnetic flux showed a sharp onset of IM response as the frequency sweeps through the resonance. A second onset at higher frequency was also observed, creating a prominent gap in the IM response. By extending those measurements to nonzero dc flux, new dynamics are revealed, including: dc flux tunabililty of the aforementioned gaps, and enhanced IM response near geometric resonance of the rf-SQUID.  These features observed experimentally are understood and analyzed theoretically through a combination of a steady state analytical modeling, and a full numerical treatment of the rf SQUID dynamics. The latter, in addition, predicts the presence of chaos in narrow parameter regimes. The understanding of intermodulation in rf-SQUID metamaterials is important for producing low-noise amplification of microwave signals and tunable filters.
\end{abstract}

\maketitle

\section{Introduction} 
Metamaterials were originally developed as a means to create novel interactions between matter and electromagnetic (EM) radiation.\cite{Ves68,Pendry96,Pendry99,Shel01}  It became clear that these novel interactions are particularly stark and dramatic in the case of metamaterials composed of superconducting 'meta-atoms'.\cite{Ricci05}  Superconductors offer low loss and the ability to create compact meta-atoms due to their high critical current densities,\cite{Anlage11} thus ensuring that the metamaterial limit constraint (meta-atom size much less than EM wavelength) is satisfied.\cite{Kur10,Kur11}  In addition, superconductors bring both macroscopic and microscopic \textit{quantum} effects into the largely classical field of metamaterial research.\cite{Jung14}  The macroscopic quantum effects include magnetic flux quantization, and the Josephson effects.  These two macroscopic properties are elegantly combined in a single meta-atom known as a radio frequency superconducting quantum interference device (rf-SQUID).  rf-SQUIDs are free-standing superconducting loops interrupted by a single Josephson junction.\cite{Silver67}  As such, they are essentially split-ring resonators in which the capacitor is replaced with a Josephson junction, to create a unique meta-atom.  The junction essentially acts as a parallel combination of a fixed capacitance, a fixed resistance, and an ideal nonlinear inductor that can be tuned to explore both positive and negative values of inductance, including $\pm \infty$.  The combination of geometrical and Josephson inductance, along with geometrical and Josephson capacitance, creates a self-resonant object that can oscillate with low loss in the microwave regime.  This simple device enables a nonlinear and tunable meta-atom, thus establishing an entirely new class of metamaterials.\cite{DuChenLi06,Laz07}  As an added benefit, rf SQUID metamaterials can have strong and long range interactions between the meta-atoms, which creates collective responses that can dramatically alter their interactions with EM fields.  These interactions are also strongly nonlinear, making rf SQUID metamaterials a unique platform for the study of nonlinear physics involving many degrees of freedom.  A number of review articles have appeared on superconducting metamaterials, and the realization of quantum effects in these unique engineered materials.\cite{Anlage11,Jung14,Laz18}

The individual rf SQUID meta-atoms, and metamaterials made from these meta-atoms, have demonstrated a number of remarkable properties.   The individual rf SQUIDs demonstrate extraordinarily broad tuning of their resonant frequency with both dc and rf magnetic flux.\cite{Jung13,Butz13,Trep13}  The dc flux tunability is periodic, repeating every time the rf SQUID is punctuated by an integer number of flux quanta $\Phi_0=h/2e$ where $h$ is Planck's constant and $e$ is the electronic charge.  This tunability, accompanied by the intrinsic nonlinearity of the Josephson effect, leads to  bistability\cite{Laz13,Muller19} and multistability\cite{Jung14a,Tsir14} in their response to rf and dc driving fields.  This in turn leads to complex and hysteretic behavior, including the phenomenon of transparency.\cite{Zhang15} Theory predicts that, under appropriate circumstances, driven rf SQUIDs will display strange nonchaotic attractors\cite{Zhou92} and chaos.\cite{Hiz18,Shena20}

Gathering the rf SQUIDs into a metamaterial can create a number of interesting properties.  These center around the question of whether a large array of interacting nominally identical rf SQUIDs will oscillate coherently, or incoherently, when driven by a uniform global rf and dc flux.  For example a finite size array will show strong edge effects because SQUIDs in the center and edges of the metamaterial will experience very different local magnetic environments.\cite{TrepT15}	Despite the disorder created by edges and defective SQUIDs, one can still observe coherent oscillations under strong rf drive.\cite{Trep17}  Alternatively, complex spatial patterns created by disorder and inhomogeneous excitation with rf fields have been predicted,\cite{Laz13,Laz17,Laz18e} and visualized by laser scanning microscopy.\cite{Jung13c,Zhu19}  One of the most surprising spatio-temporal patterns concerns chimera states of the metamaterial.  In this case a metamaterial made up of identical meta-atoms under uniform excitation will spontaneously break into domains of coherent oscillation intermingled with incoherent behavior.\cite{Laz15,Hiz16c,Hiz16s,Ban18,Hiz19,Hiz20a,Hiz20b}

rf SQUIDs and rf SQUID metamaterials have been considered for applications as well.  Examples include parametric amplification,\cite{Cast07,Cast08,Mack15,Kis19} and the demonstration of the dynamical Casimir effect in a Josephson metamaterial.\cite{Laht13}  Other applications include tunable impedance matching,\cite{Alt13} and a tunable filter.\cite{Kim19}

With more sophisticated engineering, SQUID meta-atoms can be reduced further in size  to the limit where the microscopic quantum theory applies, thus becoming qubits.  In this case the qubits have a transition between their ground state and first excited state that corresponds to the absorption or emission of a single microwave photon.  A collection of these qubits, interacting either directly or indirectly through a common microwave resonator, can create a quantum metamaterial with new collective properties.\cite{DuChenLi06,Macha14}  Predicted properties include a super-radiant state in which all of the qubits start in the excited state and are triggered to emit in phase when a single photon passes by.\cite{Ivic16}  A recent experimental result is the development of a topological edge state in a one-dimensional superconducting quantum metamaterial.\cite{Besedin21}

Superconducting metamaterials based on the Josephson effect are not limited to the SQUID architecture. In particular, a novel Josephson dielectric metamaterial with an electronically tunable plasma frequency has been demonstrated.\cite{Trep19}  This design could be useful for a tunable plasmonic haloscope to detect candidate axionic dark matter particles.\cite{Lawson19}

All of the superconducting metamaterials discussed above display strong signatures of nonlinearity,\cite{Ric06,Ric07,Kur11a,Kur12,Kur15} even to the single-photon limit.  Here we are interested in characterizing and understanding the nonlinear properties of individual rf SQUIDs, which can help inform the behavior of metamaterials made up of such objects.  In particular, we look at the case of two equal-amplitude input microwave tones and the generation of third-order intermodulation products at nearby frequencies.  Intermodulation distortion (IMD, also called IM below) arises from nonlinear properties of the rf-SQUIDs, resulting in a mixing of the two input frequencies.\cite{Zhang16}  IMD generation has been widely used to characterize nonlinear systems,\cite{Hut10,Tho11} and superconducting materials \cite{Sam95,Vop00,Oates04,Boo05,Mir09,Zhur10} and devices.\cite{Muh93,Dahm96,Sol96,Hu99,Zhur03,Buks09,Rem10}  When two tones at frequencies $f_1$ and $f_2$ are applied at a frequency separation $\Delta f \equiv |f_2-f_1| \ll f_1, f_2$, third-order tones appear at frequencies $\Delta f$ above and below the two input tones.  Measuring such tones with a spectrum analyzer is relatively simple, and the dependence of their amplitude on the microscopic properties of the SQUIDs and metamaterial structure directly probes the nonlinear Josephson physics of these materials.  Intermodulation is also of concern in communications applications because mixing of two or more tones (in an amplifier or filter, for example) can create new tones in the bandwidth of the receiver chain and produce false signals that mimic the presence of additional users.  The use of two unequal tones applied to rf SQUID metamaterials makes use of the parametric amplification effect to transfer energy from a pump tone to a signal tone.  Such experiments have been very successful at producing low-noise amplification of microwave signals.\cite{Kis19,Aum20}  These conditions are very different from those utilized here, and will not be considered further.  Alternatively, one could also measure harmonic generation under single-tone input.  However, the properties of the metamaterial, and the microwave apparatus, are very different at the harmonic frequencies, making such measurements difficult to interpret.

Previous work on two-equal-amplitude-tone excitation of rf SQUIDs concentrated on how the third-order IMD amplitude depends on temperature, the rf flux drive amplitude, and the center frequency of the two driving tones.\cite{Zhang16} That study showed a dramatic onset of IMD as a function of increasing center frequency with a strong response peak, followed by a significant dip. A second peak also occurred under high applied rf flux, creating a gap in the IM response. The sharp onset was associated with the resonant response of the rf SQUID meta-atom, and the resonance tunes over a considerable frequency range as the rf flux amplitude is varied.  The dip feature was found to be asymmetric between the upper and lower third-order intermodulation tones.  Several approximate models of IMD generation from an rf SQUID were developed and showed good agreement with experimental data.  However, all of that work was done in the limit of zero dc magnetic flux, which considerably simplified the problem, both experimentally and theoretically.\cite{Zhang16}  Consequently, several key physical phenomena went undiscovered in those studies, and that situation is rectified here.  Our objective in this paper is to extend the two-tone intermodulation study to the case of non-zero dc flux.  A number of new phenomena come into play in the presence of a dc flux that were not anticipated on the basis of prior work.\cite{Zhang16}  These phenomena are explored experimentally, and the key physics underlying their origin is uncovered through a combination of analytical and numerical investigations.

\section{Theory/Model}  
\begin{figure*}[ht]
\includegraphics[width=0.5\textwidth]{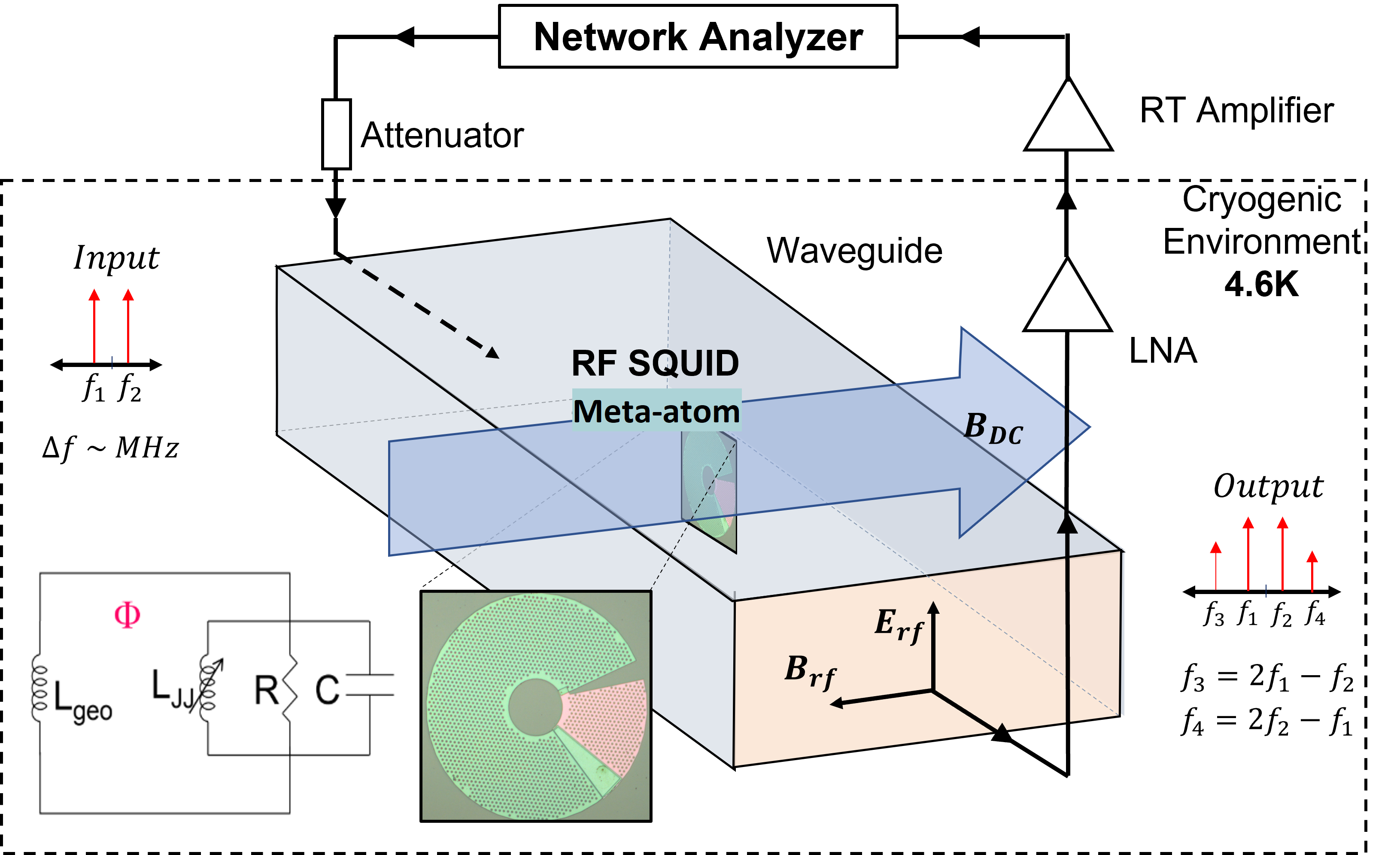}
\caption{Schematic of the experimental setup for intermodulation measurements of single rf SQUID meta-atoms.  A microwave network analyzer at room temperature generates two equal amplitude signals and sends them to a waveguide inside a cryogenic refrigerator at 4.6 K.  The signals interact with the rf SQUID meta-atom (by applying an rf magnetic flux) and produce intermodulation products which are amplified with a cryogenic low-noise amplifier (LNA) and room temperature (RT) amplifier, before returning to the network analyzer for measurement.  A perpendicular dc magnetic field can also be introduced onto the meta-atom (creating a dc magnetic flux bias) through a superconducting coil attached to the normal metal waveguide.  Lower left insets show an electrical schematic of the rf SQUID, as well as an optical micrograph of one such SQUID.  Upper left and lower right insets show the naming convention for the two input tones and the four output tones that are subsequently measured.}
\label{Expfig}
\end{figure*}

All observable properties of an rf SQUID can be predicted once the gauge-invariant phase on the Josephson junction $\delta(t)$ is known as a function of time $t$ for a given driving configuration.  For example, the supercurrent through the junction is given by $I=I_c \sin[\delta(t)]$, where $I_c$ is the critical current of the Josephson junction, and the voltage drop across the junction is given by $V=2\pi \Phi_0 d\delta/dt$.  A single rf SQUID can be treated as a resistively and capacitively shunted Josephson junction (RCSJ model) in parallel with the superconducting loop geometric inductance $L_{geo}$ (see the circuit model in Fig. \ref{Expfig}). A resistance $R$ represents dissipative quasi-particle tunneling through the junction.  Note that in this experiment the rf SQUID is designed to be nonhysteretic ($\beta_{rf} = 2\pi L_{geo} I_c/\Phi_0 < 1$).

The time evolution of the gauge invariant phase for a driven rf-SQUID described by the circuit model shown in Fig. \ref{Expfig} can be reduced to the following dimensionless form,\cite{Trep13,Zhang16,Trep17}
\begin{align}
    \label{RCSJeq}
    \frac{d^2\delta}{d\tau^2}+\frac{1}{Q}\frac{d\delta}{d\tau}+\delta+\beta_{rf} \sin\delta=2\pi(f_{dc}+f_{rf}(\tau)),
\end{align}
where $\tau=\omega_{geo}t$, $\omega_{geo}=2\pi f_{geo}=1/\sqrt{L_{geo}C}$ is the geometric resonant angular frequency of the rf SQUID in the absence of the Josephson effect, $Q=R\sqrt{C/L_{geo}}$ is the rf SQUID quality factor,  $f_{dc}=\Phi_{dc}/\Phi_0$, and $f_{rf}=\Phi_{rf}/\Phi_0$ are the normalized dc and rf magnetic fluxes. Note that the dimensionless parameter $\beta_{rf}$ controls the strength of the nonlinearity.

The two-tone excitation of the rf SQUID can be written as $f_{rf}(\tau)=f_{rf,1}\sin[\Omega_1 \tau+\theta_1]+f_{rf,2}\sin[\Omega_2 \tau+\theta_2]$, where $\Omega_{1,2}= f_{1,2}/f_{geo}$, and $f_1$ and $f_2$ are the linear frequencies of the two injected tones. Here we consider the general case in which the amplitudes and phases of the driving tones are different.  The dimensionless driving flux is more conveniently expressed in terms of a complex phasor envelope modulated by the central driving frequency,\cite{Zhang16}
\begin{align}
    \label{TwoToneEnvFunc}
    f_{rf}(\tau)=Re[e^{i\Omega\tau-i\pi/2}\tilde{f}_{rf}(\tau)],
\end{align}
where the the complex envelope function
$\tilde{f}_{rf}(\tau)=f_{rf,1}exp(-i\Delta\Omega\tau/2+i\theta_1)+f_{rf,2}exp(+i\Delta\Omega\tau/2+i\theta_2)$, the central driving frequency $\Omega=(\Omega_1+\Omega_2)/2$, and the difference frequency $\Delta\Omega=\Omega_2-\Omega_1>0$. In the case $\Delta\Omega\ll\Omega$, the relative phase between central and difference frequency components does not affect the solution. Thus we may disregard $\theta_1$ and $\theta_2$ by shifting the relative times of the central and difference frequency components.\cite{Zhang16}

Next we will discuss the steady state analytical model for the approximate solution for $\delta(\tau)$ under two-tone driving where the envelope varies so slowly that its derivatives vanish.\cite{Zhang16} This model begins with the formulation of an ansatz solution,
\begin{align}
    \label{SS_ansatz}
    \delta(\tau) = \bar{\delta} + \tilde{\delta}\sin{(\Omega\tau+\theta)}
\end{align}
where $\bar{\delta}$ is the quasi-dc offset that will change with the slowly varying envelope $\tilde{f}_{rf}$, $\tilde{\delta}$ refers to the envelope of $\delta(\tau)$, and $\theta$ is the phase of $\delta(\tau)$. In general, all three variables $\bar{\delta}$, $\tilde{\delta}$, and $\theta$ have parametric time dependence through the slowly varying envelope of the rf drive at the difference frequency $\Delta\Omega$. In substituting the ansatz Eq.(\ref{SS_ansatz}) into Eq.(\ref{RCSJeq}), the nonlinear term produces $\sin[\bar{\delta} +\tilde{\delta}\sin{(\Omega\tau+\theta)}]$ which is expanded into harmonic orders ($n\Omega$ for integer $n$) using the Jacobi-Anger expansion. A key approximation is that the higher order harmonics $n > 1$ are excluded since they are substantially suppressed by the second derivative term in Eq.(\ref{RCSJeq})). This results in $\sin\delta\approx\sin\bar{\delta}J_0(\tilde{\delta}) + 2\cos\bar{\delta}J_1(\tilde{\delta})\sin(\Omega\tau+\theta)$ where $J_0$ and $J_1$ are the Bessel functions. The equation can be separated into three coupled equations for three unknowns ($\bar{\delta}$, $\tilde{\delta}$, and $\theta$),\cite{Zhang16} 
\begin{align}
    \label{SS_coupled_1}
    (1-\Omega^2)\tilde{\delta}+2\beta_{rf}\cos\bar{\delta}J_1(\tilde{\delta})=2\pi\tilde{f}_{rf}\cos\theta,
\end{align}
\begin{align}
    \label{SS_coupled_2}
    \frac{\Omega}{Q}\tilde{\delta}=-2\pi\tilde{f}_{rf}\sin\theta,
\end{align}
\begin{align}
    \label{SS_coupled_3}
    \bar{\delta}+\beta_{rf}\sin\bar{\delta}J_0(\tilde{\delta})=2\pi f_{dc}.
\end{align}
Finally, $\delta(\tau)$ can then be constructed by solving Eqs.(\ref{SS_coupled_1})-(\ref{SS_coupled_3}) for $\bar{\delta}$, $\tilde{\delta}$, and $\theta$ for given driving flux $\tilde{f}_{rf}$ and $f_{dc}$.\cite{Zhang16}. Looking forward, this model is crucial in interpreting the IMD resonant response, and accurately predicts a bifurcation of the resonance frequency tuning curve with dc flux, $f_{dc}$. It also aids in explaining the strong generation of IMD products near geometric resonance, $\omega_{geo}$.  These experimentally observed features will be presented below.
	
\section{Experiment}
The experiments are carried out utilizing single rf-SQUID meta-atoms.  The meta-atoms were fabricated using the Hypres $0.3 \mu A/\mu m^2$ $Nb/AlO_x/Nb$ trilayer junction process on silicon substrates, and the meta-atom has a superconducting transition temperature $T_c = 9.2\ K$.\cite{Trep13}  An electrical schematic and optical micrograph of the rf SQUID are shown in Fig. \ref{Expfig}. The inner radius of the rf-SQUID sample is 100 $\mu  m$ with a geometrical area $A_0$ = 31,416 $\mu m^2$.  The effective area is estimated to increase by a factor 1.1 in comparison to the geometrical area of the rf SQUID loop due to flux focusing created by the surrounding superconducting ring ($A$ = 1.1$A_0$). The two Nb films (135 and 300 $nm$ thick) are connected by means of a via and a Josephson junction to create a superconducting loop with geometrical inductance $L_{geo}$. The capacitance $C$ has two parts: the overlap between two layers of Nb with 200 $nm$ thick $SiO_2$ dielectric in between, and the Josephson junction intrinsic capacitance.  Parameter values for a typical rf SQUID include critical current $I_c=1.15\ \mu A$, geometrical inductance $L_{geo}=  280 \ pH$, zero-bias Josephson inductance $L_{JJ,0}=\frac{\Phi_0}{2\pi I_c} = 286\ pH$, resistance $R=1780\ \Omega$, and total capacitance $C= 0.495\ pF$, geometrical resonant frequency $f_{geo}=13.52$ GHz, and $\beta_{rf}=L_{geo}/L_{JJ,0}=2\pi L_{geo} I_c/\Phi_0=0.98$, with all temperature dependent quantities taken at 4.6 K $<T_c$.

In the experimental setup shown in Fig. \ref{Expfig}, the rf SQUID sits in a normal metal Ku-band rectangular waveguide oriented so that the rf magnetic field of the propagating lowest-order TE mode is perpendicular to the rf-SQUID. This propagating mode is used to both measure the rf-SQUID response and apply a finite (and controllable) rf-flux $f_{rf}$ to the SQUID. The magnetic field of the propagating mode couples to the rf SQUID with coupling coefficient $g$ ($g \approx 0.015$), and the resulting powers from the waveguide are amplified by 55 dB, as in the experiment.\cite{Zhang16} Before each two-tone experiment, a single-tone transmission experiment is conducted to determine the resonant frequency at which the system has maximum power absorption. Intermodulation products are then measured systematically around the resonance; two signals of frequencies $f_1$ and $f_2$ having the same amplitude $f_{rf,1}=f_{rf,2}$ and a small difference in frequency $\Delta f=f_2-f_1>0$ are created by the network analyzer, attenuated, and injected into the waveguide, before reaching the sample.  We operate in the regime in which the rf flux amplitude applied to the SQUID is much less than a single flux quantum.  The output signal contains the two main tones and their harmonics, as well as intermodulation products. The set of four tones arising from the input tones and third-order IMD are measured in the network analyzer after it has been amplified by both cryogenic and room temperature amplifiers. Note that only the magnitude, but not the phase, of each of the tones is measured. Throughout this paper we shall use the notation shown in Fig. \ref{Expfig} in which the lower third-order side-band is denoted $f_3 = 2f_1-f_2$, and the upper third-order side-band is denoted $f_4 = 2f_2-f_1$.  The metamaterial exhibits strong intermodulation generation above the noise floor when superconducting (measured at $T = 4.6\ K$), and has no observable IM response at temperature above the transition temperature $T_c = 9.2\ K$.
For the results discussed in this paper, in both the experiment and simulations, these common parameters are used: $\Delta f=10\ MHz$, $\beta_{rf}=0.98$, $Q=75$, $f_{geo}=13.52$ GHz. 

\section{Numerical Simulations}

In some cases, we solve the full equation of motion for $\delta(t)$, Eq.(\ref{RCSJeq}), numerically using the LSODA solver built into the FORTRAN77 library ODEPACK. To perform scans of frequency 
or scans of rf flux amplitude 
we first find a steady state solution with initial conditions $(\delta(0),\frac{d\delta}{d\tau}(0)) = (0,0)$. The solution is considered steady once calculated over a sufficiently long time such that frequency peaks $f_1$ and $f_2$ in the Fourier transform of the time series are easily distinguished. For the parameter space examined in this paper, 20 beat periods was sufficient. In continuing the scan, the ending conditions of the previous solution seed the next solution's initial conditions. In each solution, the amplitudes $\delta_i$ of the calculated $\delta(\tau)$ corresponding to frequency components $f_i$ are extracted via Fourier transform ($i$ = 1,2,3,4). Each $\delta_{i}$ relates to the magnetic flux in the SQUID by $\delta_i = 2\pi(\Phi_i/\Phi_0)$. The magnetic flux corresponds to a magnetic field inside the single rf SQUID loop by $B_i = \Phi_i/A$ where $A$ is the effective area of loop. The final numerically simulated powers $P_{1-4}$ are calculated and presented for direct comparison to experiment.

\begin{figure*}[htp]
\includegraphics[scale=1.05]{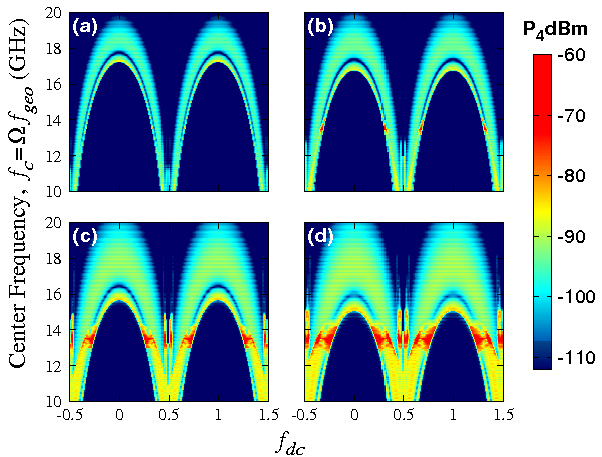}
\caption{(a)-(d) Numerical simulation data for IM product $P_4$ (colors) vs center frequency ($f_c =\Omega f_{geo}$) and dc flux ($f_{dc}$) at progressively increasing $f_{rf}$ values: (a): $f_{rf}=10^{-1.74}$ (b): $f_{rf}=10^{-1.49}$, (c): $f_{rf}=10^{-1.34}$, (d): $f_{rf}=10^{-1.24}$. Note that $f_{dc}$ values range from -0.5 to +1.5, and a common scale bar is used for the $P_4$ colors on all plots.  Additional parameters are the same as in Fig. \ref{figIMvsfanddc}.}
\label{New_Numer_Figure}
\end{figure*}

The results for the intermodulation power $P_4$, shown in Fig. \ref{New_Numer_Figure},
have been obtained through the following procedure:
For each point on the $f_{dc} - f_c$ plane, Eq. (\ref{RCSJeq}) is integrated 
in time using the standard fourth-order Runge-Kutta algorithm with fixed time-step
$h=0.02$ for $100,000$ time-units, and the data are discarded. Then, Eq. (\ref{RCSJeq})
is integrated further in time, and a time-series of $N_{FT} =2^{22}$ values of 
$\delta(\tau)$ are collected every $0.1$ time-units which span a time-interval of
$\delta \tau =419,432$ time-units. That time-interval is much larger than the longest
period of the rf SQUID dynamics, $2\pi / \Delta \Omega / 2 \simeq 16,990$, and
includes $\sim 25$ such periods. The time-series is then Fourier transformed and the
corresponding $\delta_i$ are obtained. Note that neighboring Fourier frequencies are
separated by $1/\delta \tau \simeq 2.4 \times 10^{-6}$ which allows the extraction of
$\delta_i$s with high accuracy. The intermodulation power $P_4$ are calculated and 
converted to dBm using the extracted $\delta_4$ and applying the procedure discussed 
above. Then, the intermodulation power $P_4$ in dBm is mapped onto the $f_{dc} - f_c$
plane in Fig. \ref{New_Numer_Figure} for four values of $f_{rf}$.

The dynamics of the rf SQUID can become chaotic in certain areas of the parameter spaces. 
In order to distinguish chaotic from non-chaotic states of the rf SQUID in the 
numerical simulations, we calculate the Lyapunov exponents, the most reliable
diagnostic tool for that purpose. First, Eq. (\ref{RCSJeq}) is written as a
system of three first order differential equations, where besides $\delta$ and 
$\dot{\delta}$, time is being treated also as dependent variable. Thus,
the system of three equations is autonomous, and the algorithm developed by
Wolf et al. \cite{Wolf1985} can be employed. For the autonomous first-order system,
there are three Lyapunov exponents whose sum is always $-1/Q$, a quantity that
quantifies loss in the rf SQUID, while one of them is always zero. The necessary
time-integration of the combined system of the three first-order equations and their
variational equations is performed by a Runge-Kutta fourth order algorithm with
constant time-step, typically $h=0.01$. The maximum Lyapunov exponent $\lambda_{max}$,
is then mapped on part of the $f_{dc} - f_c$ plane in Fig. \ref{Maximum_Lyapunov_Exponent}(b),
determines whether the rf SQUID is in a chaotic state ($\lambda_{max} > 0$ or not
($\lambda_{max} \leq 0$). The time allowed for the Lyapunov exponents to relax to an
almost constant value is more than $500,000$ time-units. Compared with the longest
period of the rf SQUID dynamics, $2\pi/\Delta \Omega /2 \simeq 16,990$, that
time-interval includes more than $30$ of these periods.

\section{Results}  
\begin{figure}
\includegraphics[width=0.5\textwidth]{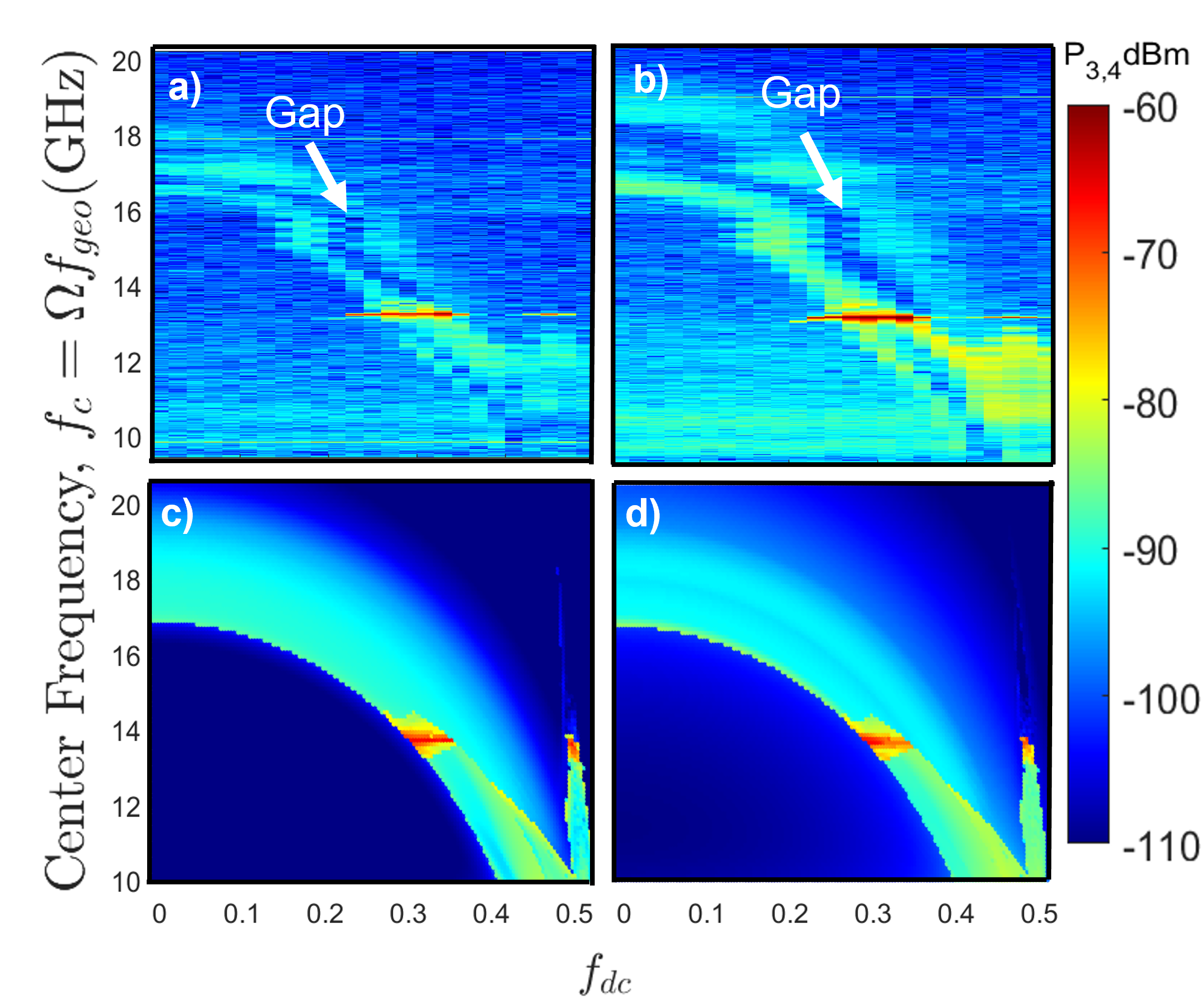}
\caption{Intermodulation powers $P_3$ and $P_4$ (colors) as a function of center frequency of the two input tones $f_c=\Omega f_{geo}=\frac{f_1+f_2}{2}$ vs.normalized dc flux $f_{dc}$ from $0$ to $\frac{1}{2}$ at normalized rf flux $f_{rf}$=$10^{-1.49}$ at temperature 4.6 K. Panels (a) and (b) show experimental data for the lower ($P_3$) and upper ($P_4$) third-order IM, respectively. Similarly, (c) and (d) show a corresponding numerical simulation for lower and upper third-order IM.  Note that the results in (a) and (b) are horizontally translated to account for trapped flux inside the rf-SQUID and/or the superconducting magnet. The coloring refers to the power of IM response in dBm as measured/simulated from a Ku band wave-guide, and a single common color bar is used for all plots. Prominent gaps in experimental IM response are pointed out in (a) and (b).  Corresponding gaps in the simulations are illustrated in Fig. \ref{figLineCuts}.  Simulation parameters: $\Delta f=10MHz$, $\beta_{rf}=0.98$, $Q=75$, $f_{geo}=13.52$ GHz.}
\label{figIMvsfanddc}
\end{figure}

\begin{figure*}
\includegraphics[scale=0.65]{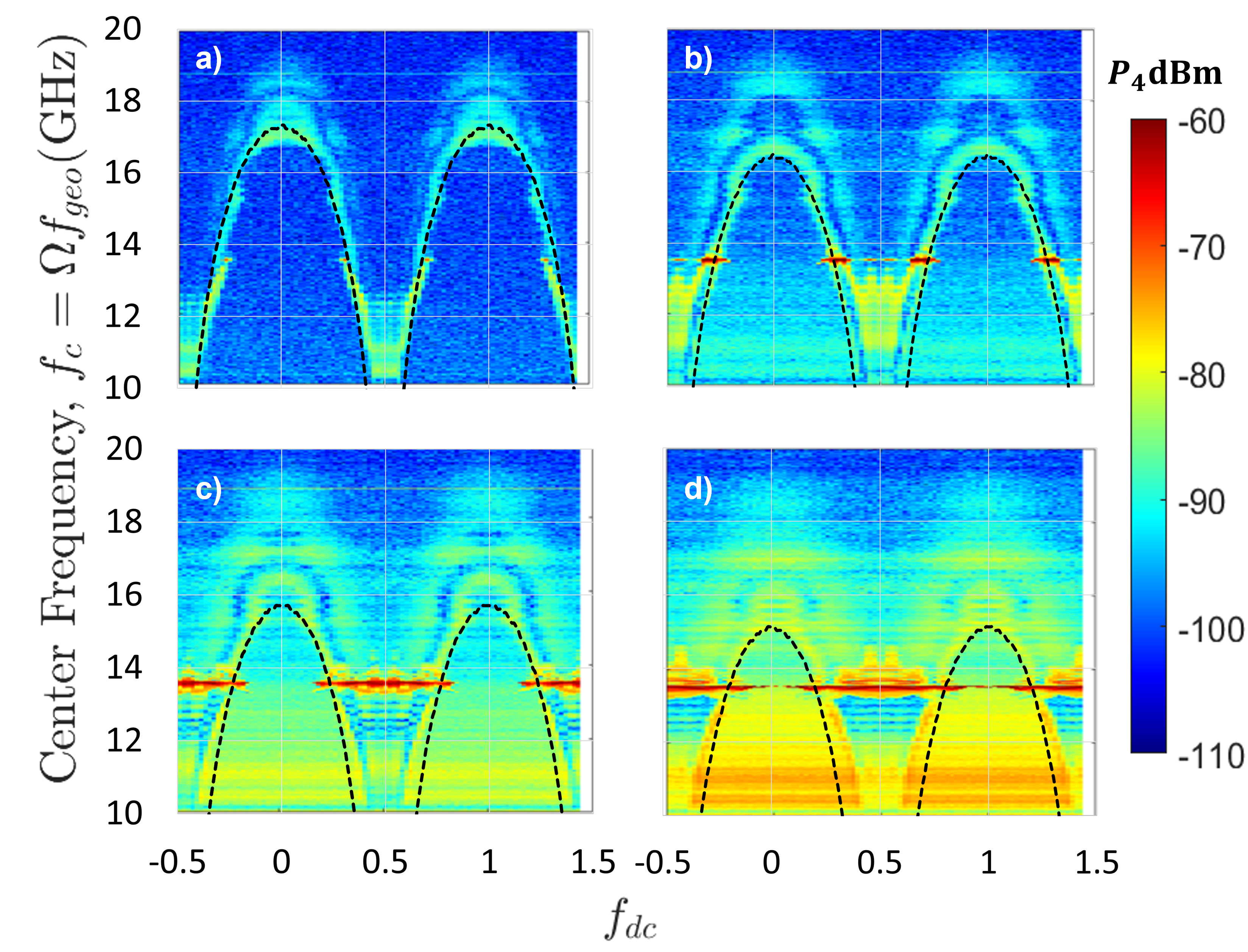}
\caption{(a)-(d) Experimental data for IM product $P_4$ (colors) vs center frequency ($\Omega f_{geo}$) and dc flux ($f_{dc}$) at progressively increasing $f_{rf}$ values: (a): $f_{rf}=10^{-1.74}$ (b): $f_{rf}=10^{-1.49}$, (c): $f_{rf}=10^{-1.34}$, (d): $f_{rf}=10^{-1.24}$. Note that $f_{dc}$ values range from -0.5 to +1.5, and the data is taken at a temperature of $T=4.6\ K$. The black dashed curves correspond to the analytically derived resonance frequency vs $f_{dc}$ evolution using the steady state model discussed in the text.  A common scale bar is used for the $P_4$ colors on all plots.  Additional parameters are the same as in Fig. \ref{figIMvsfanddc}.}
\label{figTuning}
\end{figure*}

We first give an overview of the IMD data as a function of input tone center frequency and dc flux.  Fig. \ref{figIMvsfanddc} (a) and (b) show the measured lower ($P_3$) and upper ($P_4$) third-order intermodulation strength as a function of center frequency (from 10 to 20 GHz) of the two input tones (vertical axis), and dc magnetic flux in the range of $0$ to $\Phi_0/2$ ($0\leq f_{dc} \leq \frac{1}{2}$).  We define the center driving frequency as $f_c=\Omega f_{geo}$.  For a given value of $f_{dc}$ we note that as a function of increasing center frequency there is a resonant onset of IMD, illustrated as a sharp transition from dark blue to light blue colors.  The dc magnetic flux tunes the resonant onset frequency of the rf SQUID in a periodic manner, with period $\Phi_0$ (only the first half is shown in Fig. \ref{figIMvsfanddc}, but two periods are shown in Fig. \ref{figTuning}).  This is the first demonstration of dc flux dependence of resonant response for $f_{rf}$ in the nonlinear region ($10^{-3} < f_{rf} < 10^{-1}$).  We note that the tuning of resonant frequency, and strong IMD response, is qualitatively similar to that previously seen as a function of rf flux amplitude in zero dc flux (see Fig. 2 of Ref. \cite{Zhang16}).  In Fig. \ref{figIMvsfanddc} (c) and (d), numerical simulations for the dc flux dependence of the lower and upper third-order IMD ($P_3$ and $P_4$) show similar features to the experimental data.  We note that the data and simulation results for $P_3$ and $P_4$ in Fig. \ref{figIMvsfanddc} are plotted on a common color bar, and the results show many common features.  These include the asymmetry of response between $P_3$ and $P_4$.

Figure \ref{figTuning} shows that the tuning of the $P_4$ IMD resonant response with dc flux $f_{dc}$ is expected to be periodic in units of the flux quantum $\Phi_0$.  This figure also illustrates the changes in rf-SQUID IM as the rf flux amplitude $f_{rf}$ is varied through progressively higher values in the nonlinear rf-response regime.  We note that the frequency tuning range of IM is reduced with increasing $f_{rf}$, and strong IM begins to develop near the geometrical resonance frequency of the rf-SQUID, $f_{geo}$.  Figure \ref{New_Numer_Figure} shows numerical solutions for $P_4$ for the corresponding experimental conditions shown in Fig. \ref{figTuning}.  A number of features are reproduced from the data, including the strong onset of IM generation with increasing center frequency $f_c$ at fixed dc flux $f_{dc}$, a clear gap in $P_4$ above the onset, and increased generation of IM near the geometrical resonance frequency with increased rf flux $f_{rf}$.

From these results, three dominant features characterize the IMD response: A) the resonant frequency tuning with dc flux, B) a gap in $P_4$ response above the tuning curve as a function of frequency (or dc flux), and C) a sharp enhancement of IM output power as the tuning curve crosses the geometric resonance ($f_{geo}=13.52$ GHz).  Each of these features in the experimental data will be addressed in more detail below and compared to analytical and numerical modeling.

\subsection{IMD Resonant Frequency Tuning}
\begin{figure}
\includegraphics[width=0.47\textwidth]{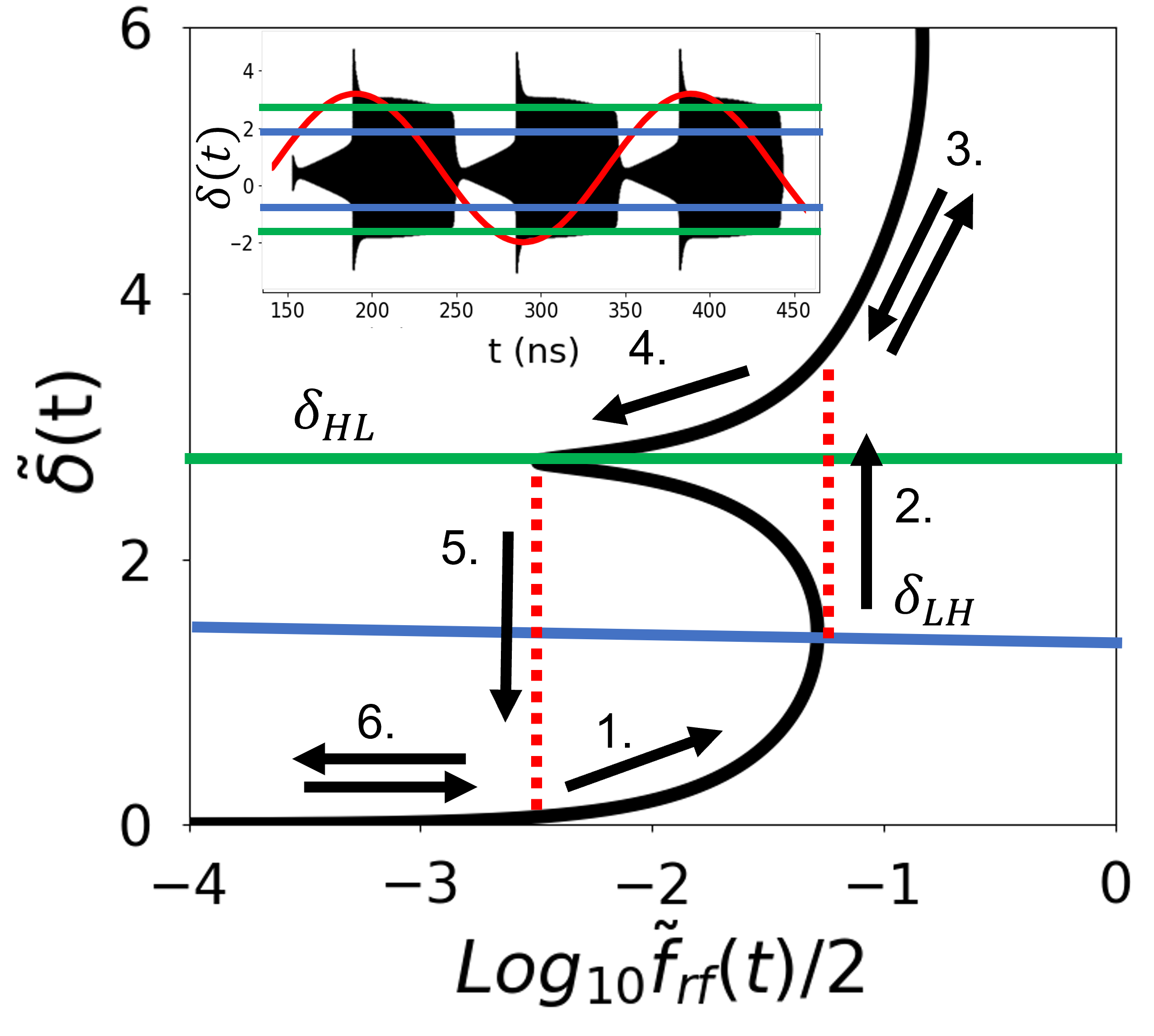}
\caption{Illustrates bistability in $\tilde{\delta}$ using the steady state model. The black curve shows a representative solution to Eqs. (\ref{SS_coupled_1})-(\ref{SS_coupled_3}) for $\tilde{\delta}$ vs. $\tilde{f}_{rf}$ for a given $f_{dc}$. The numbered arrows describe the transitions of the time-dependent envelope $\tilde{\delta}(t)$.  Inset: The resulting time dependence of $\delta(t)$ (black). The red curve displays $\tilde{f}_{rf}(t)$ scaled vertically to fit the figure.  The horizontal blue and green lines correspond to those in the main figure.}
\label{figBistable}
\end{figure}

First we will examine the tuning of the rf SQUID resonant response with respect to dc flux, at various rf flux amplitudes, as illustrated in Fig. \ref{figTuning}. In the low rf-flux limit ($f_{rf} < 10^{-3}$), the analytical linear-response model predicts that the single-SQUID resonance frequency tunes from $f_{geo}\sqrt{1 + \beta_{rf}}$ when $f_{dc} = n$ to $f_{geo}\sqrt{1 - \beta_{rf}}$ when $f_{dc} = n/2$, for all positive and negative integers $n$, including zero (here we assume $\beta_{rf}<1$).\cite{Zhang15} Using the steady state analytical model (outlined above), this concept of resonance can be extended to the IM response. In this framework, one can define IM resonance in terms of the bistable transition in the envelope of the gauge invariant phase difference $\tilde{\delta}(\tau)$.  The model shows that there is a large increase in $P_3$ and $P_4$ when $\tilde{\delta}(\tau)$ undergoes a bistable transition.  In solving the system of equations Eqs. (\ref{SS_coupled_1})-(\ref{SS_coupled_3}) using a constant $f_{dc}$, a solution curve for $\tilde{\delta}$ vs $\tilde{f}_{rf}$ is constructed, as illustrated in Fig. \ref{figBistable}. The thick black curve relates the envelope of the gauge invariant phase difference $\tilde{\delta}$ to the envelope of the two-tone driving rf-flux $\tilde{f}_{rf}$, which are both time dependant on the scale of the difference frequency ($\Delta\Omega$). (Note that the horizontal axis of this plot is on log scale of $\tilde{f}_{rf}/2$. Since the rf drive consists of two equal-amplitude tones, $\tilde{f}_{rf}/2$ thus corresponds to the power of one of the two tones.)

Bistable transitions in $\tilde{\delta}$ can be described by a cycle over the course of half a beat period in $\tilde{f}_{rf}$. In reference to Fig. \ref{figBistable}, let the cycle begin at $\tilde{f}_{rf} = 0$ and consider increasing $\tilde{f}_{rf}$. 1: $\tilde{\delta}$ will increase on the lowest branch of the black line as $\tilde{f}_{rf}$ increases until $\tilde{\delta}$ is equal to $\delta_{LH}$ (horizontal blue line in Fig. \ref{figBistable}). 2: $\tilde{\delta}$ is faced with a crisis and transitions to the next higher branch. 3: $\tilde{\delta}$ will increase on this upper branch until $\tilde{f}_{rf}$ reaches its maximum value. 4: $\tilde{\delta}$ will decrease on the higher branch as $\tilde{f}_{rf}$ decreases until $\tilde{\delta}$ is equal to $\delta_{HL}$ (horizontal green line in Fig. \ref{figBistable}). 5: $\tilde{\delta}$ is faced with another crisis and transitions to the lower branch. 6: $\tilde{\delta}$ continues to decrease as $\tilde{f}_{rf}$ decreases until passing $\tilde{f}_{rf} = 0$ and repeating the cycle for negative $\tilde{f}_{rf}$. This results in two (hysteretic) transitions in $\tilde{\delta}$ per cycle of $\tilde{f}_{rf}$ as seen in the inset plot of $\delta(t)$ in Fig. \ref{figBistable}. In reference to this cycle, we define IMD resonance as the point where the maximum of $\tilde{f}_{rf}$ coincides with the initial crisis transition such that step 3 begins to occur. In Fig. \ref{figBistable} this would occur at $log_{10}(\tilde{f}_{rf}/2) \approx -1.2$. 

The inset in Fig. \ref{figBistable} shows the resulting waveform of $\delta(t)$, in which the bistable transitions in the envelope  $\tilde{\delta}(t)$ results in large IMD production. 
Note that these curves may have several crisis locations resulting in multiple bistable transitions over one half beat of rf flux, and result in prolific production of IMD.

Analytically derived tuning curves in $f_c$ vs $f_{dc}$ can be constructed by increasing the frequency used in the steady state analytical model until the condition for resonance to occur (described above) is satisfied. The frequency at which this condition occurs for a given $f_{rf}$ and $f_{dc}$ is defined as the IM resonant frequency. The resonance frequency can be determined for varying $f_{dc}$, resulting in tuning curves in $f_c$ vs $f_{dc}$. Figure \ref{figTuning}(a)-(d) shows these analytically derived curves as dashed lines which are superimposed on experimental results taken at corresponding $f_{rf}$ levels. Note that the IM resonance in this context corresponds to the onset of strong IM response from the transitions between different solution branches, rather than a peak in the response. Thus, strong IM response should be found near but not exactly at this resonance. There is good agreement between the experimental data and model with regards to IMD resonant frequency tuning. Specifically, as $f_{rf}$ increases, the frequency range of IMD resonance tuning by means of dc flux is reduced. This is most apparent when observing how the resonant frequency at $f_{dc} = n$ approaches $f_{geo}$ as $f_{rf}$ increases.  To further solidify this point we utilize full numerical solutions to the rf SQUID equation of motion, Eq.(\ref{RCSJeq}), under the same conditions as the data in Fig. \ref{figTuning}, and the results are shown in Fig. \ref{New_Numer_Figure}.  We find excellent agreement between the experimental and the numerical results in terms of the dc flux tuning of IMD resonance with varying $f_{rf}$.

\subsection{Gaps in IMD Power}
\begin{figure}
\includegraphics[width=0.47\textwidth]{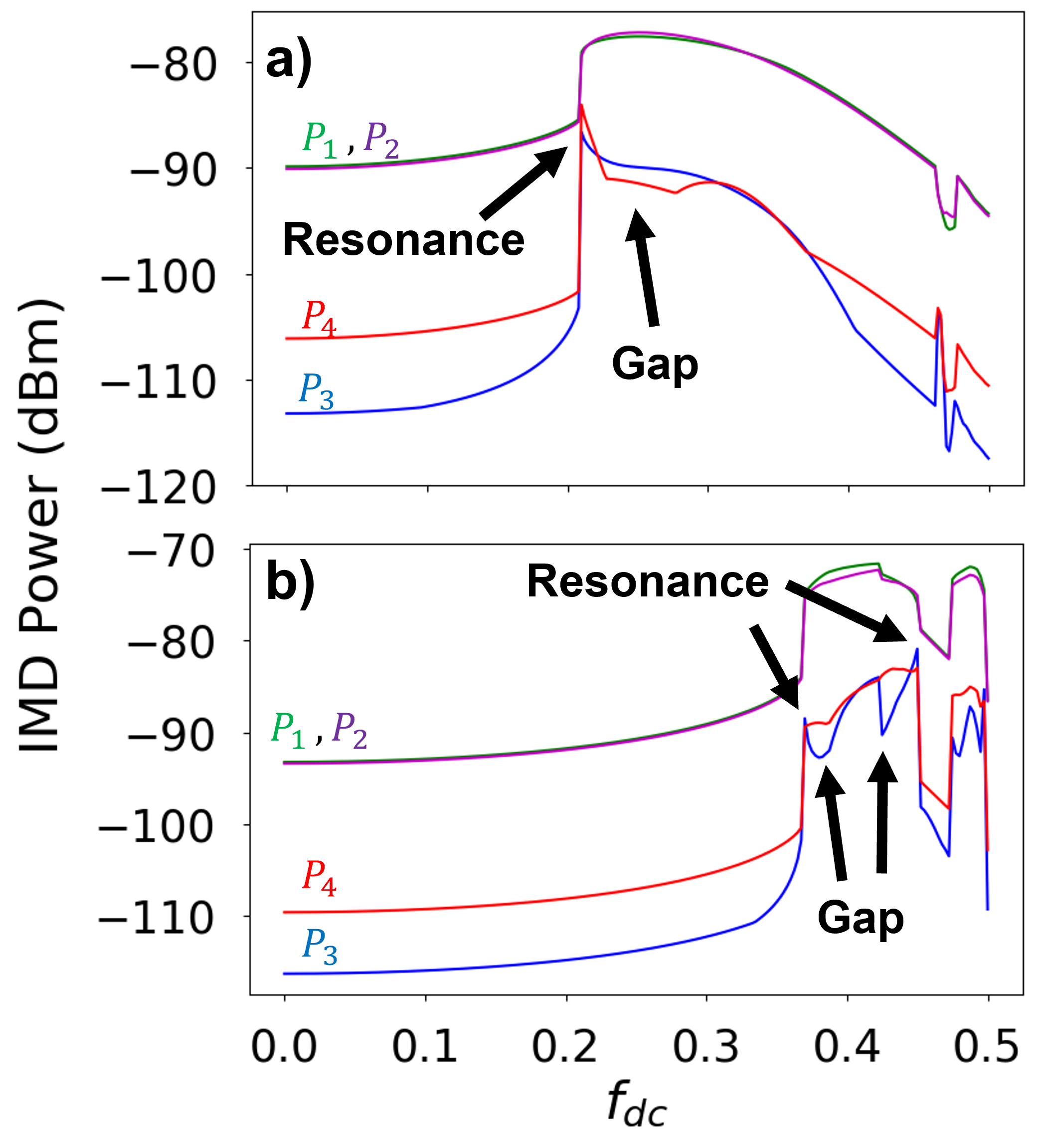}
\caption{Fundamental input ($P_1$ and $P_2$) and IM output ($P_3$ and $P_4$) power (dBm) vs dc flux $f_{dc}$, illustrating gaps in both lower and upper third-order IM.  Panels (a) and (b) are horizontal constant frequency line cuts from numerical simulations presented in Fig. \ref{figIMvsfanddc}(c) and (d) at constant center driving frequency $f_c=1.11 f_{geo} = 15$ GHz and $f_c=0.81 f_{geo}=11$ GHz, respectively.}
\label{figLineCuts}
\end{figure}

We next discuss the IMD gap above resonance evident in the data in Figs. \ref{figIMvsfanddc} and \ref{figTuning}. In the zero dc flux limit, a gap in IMD was observed above the resonance frequency.\cite{Zhang16}  Figures \ref{figIMvsfanddc} and \ref{figTuning} show that this gap continuously tracks with the resonance frequency tuning curve, creating a segment of low IMD response.  The numerical results for $P_4(f_c, f_{dc})$ in Fig. \ref{New_Numer_Figure} parallel the data shown in Fig. \ref{figTuning}.  Here the gaps are quite prominent, and display a different character above and below the geometric resonance frequency $f_{geo}=13.52$ GHz, similar to the data.  Figure \ref{figLineCuts} shows these gaps explicitly by taking horizontal (constant frequency $f_c$) cuts of numerical results in Fig. \ref{figIMvsfanddc}(c) and (d). The gap traces out the tuning curve such that it exists in both the frequency ($f_c$) and dc flux ($f_{dc}$) domains.  Figure \ref{figLineCuts}(a) shows this gap above resonance in dc flux for a constant center frequency $f_c=1.11 f_{geo}=15$ GHz while Fig. \ref{figLineCuts}(b) is a line cut taken at $f_c=0.88 f_{geo}=12$ GHz.  Within each gap, there is an asymmetry between $P_3$ and $P_4$: when $f > f_{geo}, P_4 < P_3$ and when $f < f_{geo}, P_4 > P_3$. In addition to the gap immediately following resonance in Fig. \ref{figLineCuts}(b), there is a secondary gap at larger dc flux $f_{dc} \approx 0.42$. The gap is particularly clear in the numerical simulation shown in Fig. \ref{New_Numer_Figure}.  As $f_{rf}$ increases to the nonlinear regime, a gap below $f_{geo}$ develops and becomes more distinct. This is in qualitative agreement with data shown in Fig. \ref{figTuning}.  Further details about the origin and asymmetry of the gaps in $P_3$ and $P_4$ are discussed in our previous work.\cite{Zhang16,ZhangT16} 

\begin{figure*}
\includegraphics[width=\textwidth]{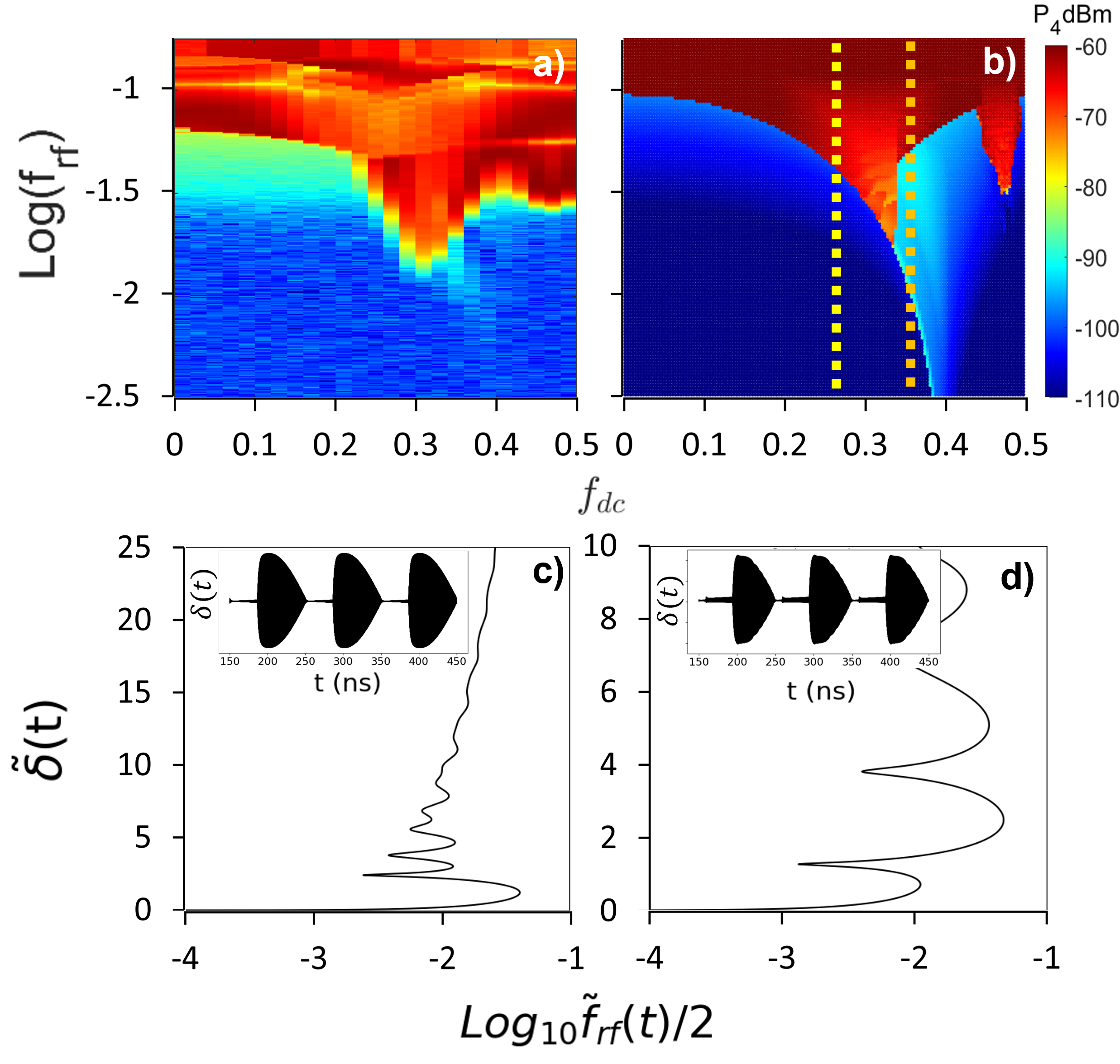}
\caption{Panels (a) and (b) show experimental data and numerical simulation, respectively, for the upper third IM power $P_4$ (shown on a common color bar in dBm) as a function of $f_{rf}$ and $f_{dc}$, taken at $f_c=0.999 f_{geo}=13.50$ GHz while sweeping $f_{rf}$ from low to high. (c) Shows the solutions to Eqs. (\ref{SS_coupled_1})-(\ref{SS_coupled_3}) for $\tilde{\delta}$ vs. $\tilde{f}_{rf}$ at $f_{dc}$=0.25 when $f_{rf}\approx10^{-1.3}$.  This case corresponds to the left vertical yellow line in (b). Inset: $\delta(t)$ at $f_{rf}=10^{-1.3}$. (d) Shows the solutions to Eqs. (\ref{SS_coupled_1})-(\ref{SS_coupled_3}) at $f_{dc}$=0.35 when $f_{rf}\approx10^{-1.3}$. This case corresponds to the right vertical orange line in (b).  Inset: $\delta(t)$ at $f_{rf}=10^{-1.3}$.}
\label{figPhirfvsPhidc}
\end{figure*}

\subsection{Enhancement of IMD near the Geometric Resonant Frequency $f_{geo}$}
It is clear from the data in Figs. \ref{figIMvsfanddc} and \ref{figTuning} that for certain ranges of $f_{dc}$ there is enhanced IMD response for drive center frequencies $f_c$ near $f_{geo}= 13.52$ GHz. This enhancement is hardly surprising  since extreme multistability near $f_{geo}$ has been modeled and observed before albeit at $f_{dc}=0$.\cite{Jung14a}  To investigate further, we examine the strong enhancement in IMD power for $f_c$ close to $f_{geo}$, as a function of $f_{rf}$ and $f_{dc}$, as shown in Fig. \ref{figPhirfvsPhidc}(a) (data) and (b) (simulation). In this case we choose a center frequency $f_c=0.998 f_{geo}=13.49$ GHz, just below the geometric resonant frequency.  Little difference was observed between $P_3$ and $P_4$ in regards to the following discussion, so only $P_4$ is considered. In addition, there was minimal hysteresis in rf flux sweep with two-equal-amplitude stimulation.\cite{Zhang16} Thus, only the low-to-high rf flux sweeps are shown. There are three common features worth noting between the experiment and the simulation shown in Fig. \ref{figPhirfvsPhidc}. 

The first feature is the main IMD resonance tuning curve, which extends from approximately $f_{rf} = 10^{-1}$ at $f_{dc} = 0$ to $f_{rf} < 10^{-2.5}$ at $f_{dc} = 0.4$ in Fig. \ref{figPhirfvsPhidc}(b). This curve represents the IMD resonance condition in $f_{rf}-f_{dc}$ space for $f_c=0.998 f_{geo} = 13.49$ GHz. As $f_{rf}$ is increased across this resonance curve, a massive increase in IMD production of $\approx 40\ dB$ can be observed.  Note that the data in Fig.  \ref{figPhirfvsPhidc}(a) shows evidence of the same tuning curve, with a trace of the transition continuing beneath the ``tooth'' of IMD response centered at $f_{dc} \approx 0.3$.  The mechanism underlying the appearance of this tooth will be discussed next.

The second common feature between the experiment and the simulation is the bifurcation of the IMD resonance tuning curve near $f_{dc} =0.3$. Its origin can be qualitatively understood through the steady state analytical model.  Figure \ref{figPhirfvsPhidc} (c), (d) show the $\tilde{\delta}-\tilde{f}_{rf}$ solution curves from the steady state model for two values of $f_{dc}$, before, and after, the bifurcation.  In Fig. \ref{figPhirfvsPhidc}(c) where $f_{dc}=0.25$ is below the $f_{dc}$ value for the onset of bifurcation, we can see that for any $f_{rf} \gtrsim 10^{-1.4}$ a bistable transition would occur, and this transition increases $\tilde{\delta}$ by an order of magnitude. On the other hand in Fig. \ref{figPhirfvsPhidc}(d) where $f_{dc}=0.35$ is above the value for the onset of bifurcation, two transitions may occur with one at $\tilde{f}_{rf}/2 \approx 10^{-1.9}$ corresponding to the lower resonance in Fig. \ref{figPhirfvsPhidc}(a) and (b), and another at $\tilde{f}_{rf}/2 \approx 10^{-1.3}$ corresponding to the higher resonance in Fig. \ref{figPhirfvsPhidc}(a) and (b). This is also observed in the inset diagram of Fig. \ref{figPhirfvsPhidc}(d) where the envelope of $\delta(t)$ has an initial small transition ($\Delta \tilde{\delta} \approx 1$) followed by the large secondary transition ($\Delta \tilde{\delta} \approx 40$) during a single beat. The dominating second transition also explains the much stronger IMD response above the higher resonance compared to the lower after the bifurcation. 

\begin{figure}
\includegraphics[width=0.47\textwidth]{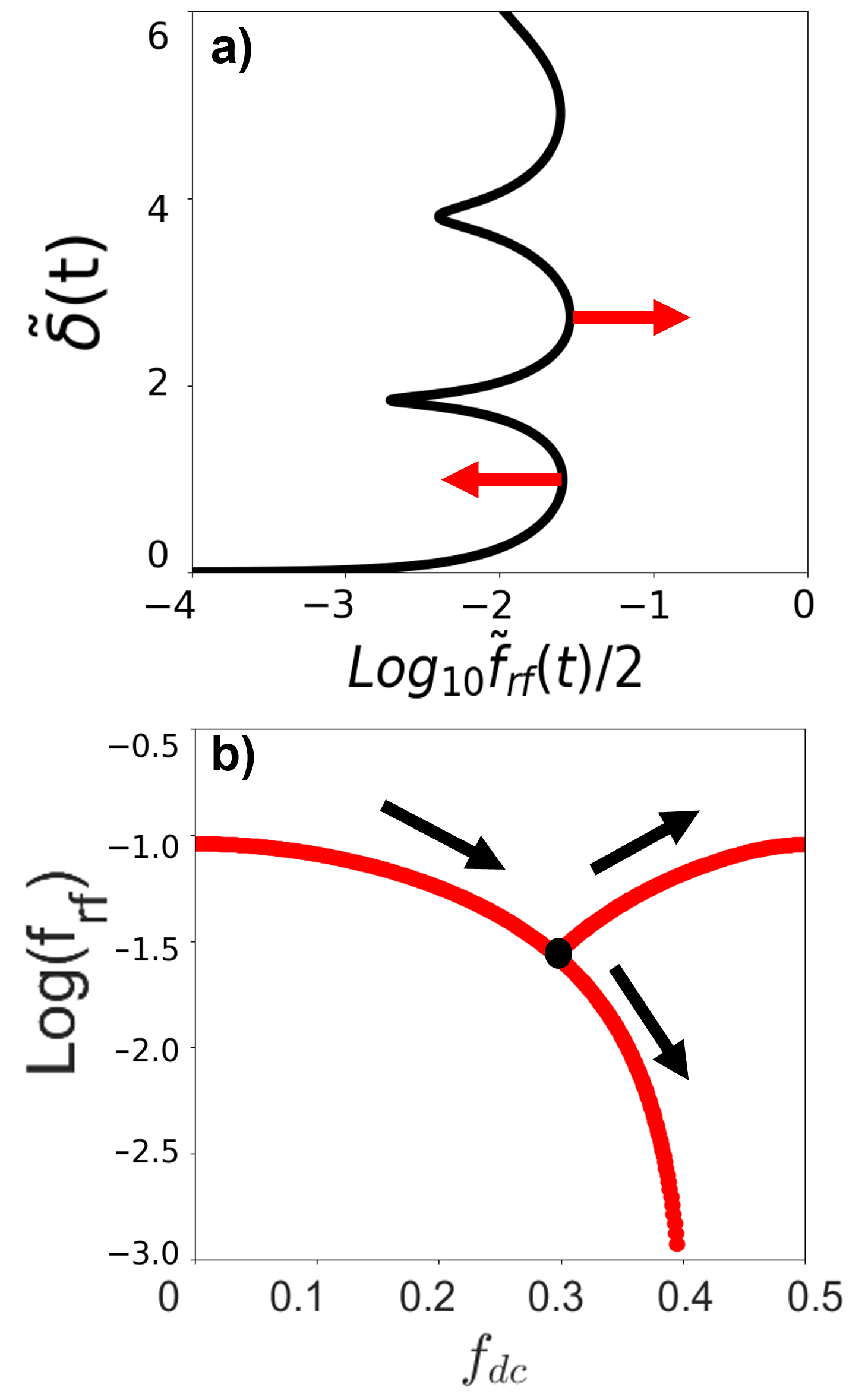}
\caption{Illustration of bifurcation of resonance frequency according to the steady state analytical model. At $f_c= 0.998 f_{geo} = 13.49$ GHz, resonance bifurcation begins at $f_{dc}$=0.3. (a) Solution for $\tilde{\delta}$ vs. $\tilde{f}_{rf}$ for $f_{dc}$=0.3. Red arrows indicate the movement of transition peaks as $f_{dc}$ increases. $f_{dc}>$ 0.3 allows two transitions below $f_{rf}$=0.1, $f_{dc}<$ 0.3 allows only one. (b) Displays the resonance in $f_{rf}$ vs $f_{dc}$ at $f_c= 0.998 f_{geo} = 13.49$ GHz. The black dot represents the point of bifurcation shown in (a).}
\label{figBifur}
\end{figure}

Figure \ref{figBifur}(a) shows the $\tilde{\delta}-\tilde{f}_{rf}$ solution curve at exactly the bifurcation point, where there are two vertically aligned crisis locations. The crisis points will move in $\tilde{f}_{rf}$ as $f_{dc}$ increases according to the arrows shown, such that for $f_{dc}>$ 0.3 two transitions may occur as in Fig. \ref{figPhirfvsPhidc}(d) , while for $f_{dc}<$ 0.3 only one transition occurs as in Fig.  \ref{figPhirfvsPhidc} (c). The resulting bifurcation of the IMD resonance frequency in $f_{rf}$ vs $f_{dc}$ can be seen in Fig. \ref{figBifur}(b), which qualitatively reproduces the main bifurcation and the ``tooth'' of high IMD response at $f_{dc}=0.32$ seen in the $f_{rf}-f_{dc}$ plots in Fig. \ref{figPhirfvsPhidc}(a) and (b).

The third common feature between the experiment and the simulation is the second ``tooth'' of high IMD response projecting into lower $f_{rf}$ located at approximately $f_{dc} = 0.48$. Unlike the other two features, this ``tooth''  cannot be explained by the steady state analytical model. It is hypothesized that the exclusion of higher order harmonics in the steady state model is the cause of this issue. 

Some connections can be made by comparing the enhancement of IMD response seen from tuning curves in different parameter spaces. Specifically, a line cut at $f_{rf}=10^{-1.49}$ in the $f_{rf}-f_{dc}$ space for a constant $f_c$ (see Fig. \ref{figPhirfvsPhidc} (a) and (b)) can be transformed into the line cut at $f_c=0.999 f_{geo}=13.50$ GHz in $f_c-f_{dc}$ for a constant $f_{rf}$ (see Fig. \ref{figIMvsfanddc}(b) and (d)). As the main tuning curve passes $f_{geo}$ in Fig. \ref{figIMvsfanddc}(b) and (d), it splits into two resonance branches just as the above-mentioned line cut in Fig. \ref{figPhirfvsPhidc} (a) and (b) intersects with both of the ``teeth'' structures. The lower branch has a much weaker response and gradually dissipates as $f_{dc}$ increases to $0.5$, while the upper branch is much stronger. This is seen more clearly in Fig. \ref{figLineCuts}(b) when there are two resonances and two gaps for $f_{dc} < 0.48$. The resonance at $f_{dc} \approx 0.46$ corresponds to the upper branch and has a stronger response than the lower at $f_{dc} \approx 0.38$. Similarly, in Fig. \ref{figPhirfvsPhidc} (a) and (b), the line cut at $f_{rf}=10^{-1.49}$ has a stronger response near the second ``tooth'' where $f_{dc}=0.48$ compared to the response at the first ``tooth'' where $f_{dc}=0.32$.
	
\section{Discussion} 
\label{sec:discussion}
There are other aspects of nonlinear dynamics that could be observed in the intermodulation measurement. Here we confine ourselves to discussing the possibility of chaos induced in the rf SQUID response by a two-tone rf drive, combined with a finite dc flux $f_{dc}$.

\begin{figure}
\includegraphics[scale=0.75]{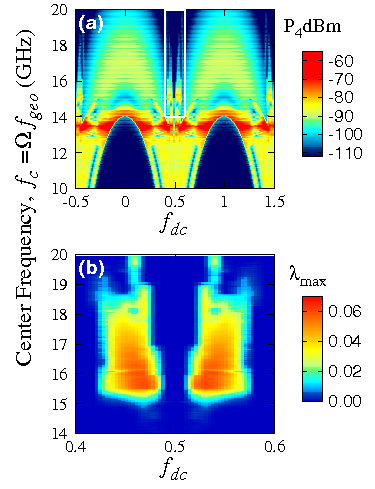}
\caption{(a) Numerical data for IM product $P_4$ (colors) vs center frequency ($f_c =\Omega f_{geo}$) and dc flux ($f_{dc}$) at $f_{rf} =10^{-1.10}$.  
(b) The corresponding maximum Lyapunov exponent shown for the area enclosed in the white rectangle in (a). 
Additional parameters are the same as in Fig. \ref{figIMvsfanddc}.}
\label{Maximum_Lyapunov_Exponent}
\end{figure}

 
The dynamics of Eq. (\ref{RCSJeq}) can be chaotic in certain regions of parameter space (see ~\cite{Hiz18} and references within).
The rf SQUID is mainly, for most of the external control parameter values, 
in a quasiperiodic state due to the two-tone quasiperiodic driving field. The distinction between the chaos and quasiperiodicity of an rf SQUID is nontrivial experimentally, but numerically possible through the calculation 
of the Lyapunov exponents. Let us consider the parameter space of $f_{dc}$ vs.
the center frequency $f_c$ as in Fig. \ref{Maximum_Lyapunov_Exponent}. Our numerical simulations have revealed that chaotic
states appear very close to half-integer bias flux $f_{dc}$, for $f_c$ greater than the geometrical frequency $f_{geo} =13.52$ GHz.
The areas on the $f_{dc} - f_c$ plane in which the rf SQUID is in a chaotic
state grows with increasing amplitude of the external rf field, $f_{rf}$. 

In Fig. \ref{Maximum_Lyapunov_Exponent}(a), intermodulation power $P_4(dBm)$, for
$f_{rf} =10^{-1.1}$ is mapped on to the $f_{dc} - f_c$ parameter plane. The observed pattern
exhibits similar characteristics as those observed in Fig.
\ref{New_Numer_Figure}. However, 
due to the relatively high value of $f_{rf}$, the rf SQUID exhibits strong
response around the geometrical frequency $f_{geo}$ (red features) over almost
the entire range of $f_{dc}$ values. Even for that relatively high value
of $f_{rf}$, however, in most of the area of the $f_{dc} - f_c$ parameter plane
the rf SQUID is in a non-chaotic state. 

In Fig. \ref{Maximum_Lyapunov_Exponent}(b), the maximum Lyapunov exponent
$\lambda_{max}$ is mapped for part of that plane around $f_{dc}=0.5$ (enclosed
in the white rectangle in (a)). In the blue areas of the map, $\lambda_{max}$ 
is zero or just below zero, indicating that the rf SQUID is in a non-chaotic
state.  In the rest of the area of the plane, $\lambda_{max}$ is greater than
zero, indicating that the rf SQUID is in a chaotic state. Note that there are two
relatively small chaotic areas in that part of the $f_{dc} - f_c$ plane, which
are located symmetrically around $f_{dc}=0.5$. Such chaotic areas, of
similar size and shape, appear around any half-integer value of $f_{dc}$ in the
same range of frequencies. The chaotic areas shrink with decreasing $f_{rf}$ until they
practically vanish for $f_{rf} \lesssim 0.025$. It should be also noted that
chaotic areas of similar shape appear on the $f_{dc} - f_{rf}$ parameter plane
which are symmetrically located around half-integer values of $f_{dc}$, for
$f_{rf} =0.05$ to $0.20$ (not shown). 
 
\section{Conclusions}
We have observed the first tuning of intermodulation response with dc flux applied to a single rf SQUID meta-atom. Above the IM resonance defined in this paper, prominent gaps are observed and also tunable with dc magnetic flux. The IM response is enhanced as the IM resonance is tuned through the geometric resonance of the SQUID. The enhanced IM response near geometric resonance also shows a bifurcation in $f_{rf}-f_{dc}$ space. All of these features are understood in a semi-quantitative manner through a combination of a steady state approximation model, and a full numerical treatment, of the rf SQUID dynamics. The numerical solutions also predict the presence of chaos in narrow parameter regimes.

\section{Acknowledgements}
The work at Maryland is funded by the US Department of Energy
through Grant $\#$DESC0018788. We acknowledge use
of facilities at the Maryland Quantum Materials Center
and the Maryland NanoCenter.
J.H. and N.L. acknowledge support by the General Secretariat 
for Research and Innovation (GSRI) and the Hellenic Foundation 
for Research and Innovation (HFRI) (Code No. 203).

\bibliography{WTD_new.bib}














\end{document}